\pdfoutput=1
\documentclass[9pt,technote]{IEEEtran}
\usepackage{amsmath,amssymb}
\usepackage{graphicx}
\usepackage{cite}
\usepackage{booktabs}
\usepackage{url}
\usepackage{xcolor}
\makeatletter
\def\maketitle{\par\begingroup\normalfont
  \def\thefootnote{}\def\footnotemark{}\let\@makefnmark\relax
  \footnotesize\footnotesep 0.7\baselineskip\normalsize
  \twocolumn[{\centering\vspace*{0pt}{\@title\par}\vskip 8pt{\@author\par}\vskip 8pt}]%
  \thispagestyle{IEEEtitlepagestyle}\@thanks
  \endgroup
  \setcounter{footnote}{0}\let\maketitle\relax\let\@maketitle\relax
  \gdef\@thanks{}\let\thanks\relax}
\makeatother

\begin{document}
\AddToHookNext{shipout/foreground}{\put(307,-770){\makebox[0pt]{\parbox{480pt}{\centering\fontsize{7}{8.4}\selectfont This work has been submitted to the IEEE for possible publication. Copyright may be transferred without notice, after which this version may no longer be accessible.}}}}

\title{\fontsize{16}{19}\selectfont\mdseries Device-Referred Channel Sounding With Commodity Wi-Fi Receivers: Error Budget and Body-Worn Measurements at 5 and 6~GHz}

\author{\fontsize{11}{13}\selectfont Raghuram C. Kamath, Carlos Aldana, and Kavya Guruswamy%
\thanks{This work involved human subjects in its research. The authors confirm that all human subject research procedures and protocols are exempt from review board approval. All participants gave informed consent.}%
\thanks{R. C. Kamath, C. Aldana, and K. Guruswamy are with Meta Platforms, Inc., Sunnyvale, CA, USA (e-mail: raghuck@meta.com; caldana@meta.com; kavyag@meta.com). (Corresponding author: Raghuram C. Kamath.)}}

\markboth{}{}
\maketitle

\begin{abstract}
\looseness=-1 A wireless product experiences the channel through its own antennas and, when worn, its user's body. Commodity Wi-Fi receivers report channel state information (CSI) per frame and can serve as device-referred sounders if their impairments are corrected. Synthetic ground truth reproducing a commercial receiver's impairments shows that five corrections are needed: subcarrier mapping, receive-chain order, per-packet alignment, received-signal-strength referencing, and window-response removal. Without them, the RMS delay spread reads up to 191\% high, the coherence bandwidth falls to about half, or the Rician $K$-factor and Doppler estimates lose their meaning; with all five, the median delay-spread bias is between $-6\%$ and $+14\%$. Applied to beacons from seven access points, including a body-worn receiver at six indoor sites at 5 and 6~GHz, the chain gives delay spreads of 24--96~ns; walking lowers the median correlation over one beacon interval from 0.85 to 0.15, near the estimator's floor. Unprecoded data frames gave 0.2--1.2 times the beacons' delay spread on the same channel, so the two should not be pooled.
\end{abstract}


\section{Introduction}
Channel sounders characterize propagation with calibrated transmitters, reference antennas, and wide bandwidth \cite{salous2013radio}. A product sees a different channel: one that includes its own antennas, its placement, and, for a wearable device, the body of its user, whose posture and motion shadow paths and change the fading \cite{hall2007antennas,cotton2009experimental,yazdandoost2008channel}, including at 5.8~GHz for moving users \cite{ferreira2024influence}. Measuring that \emph{device-referred} channel at scale calls for a sounder that is the product itself.

\looseness=-1 Commodity Wi-Fi receivers make this possible. They export per-subcarrier channel state information (CSI) for every received frame \cite{halperin2011tool,gringoli2019free,gringoli2021axcsi,jiang2022picoscenes}, which already supports a broad sensing literature and a sensing amendment to IEEE 802.11 \cite{ma2019wifi,du2025overview,ieee80211bf}. CSI has been used to recover power delay profiles by band stitching \cite{xie2015precise}, for delay-synchronous wideband sounding with a cabled reference antenna \cite{yamamoto2025delay}, for over-the-air referencing \cite{yamamoto2026radar}, and on software-defined receivers that capture CSI from commercial devices \cite{zumegen2024software}. Receiver effects such as timing and phase offsets, gain control, and filtering are documented \cite{portner2026same,jiang2022picoscenes,cominelli2023exposing,cominelli2025glimpse,tonini2025towards}; one research platform reports which antenna feeds each chain \cite{halperin2011tool}, but this client does not, which motivates correction~2. Surveys of CSI datasets also find that the processing details needed to reproduce results are often missing \cite{guarino2026survey}. What is missing is an error budget that states how much each processing step contributes to the standard channel parameters (RMS delay spread, coherence bandwidth, Rician $K$-factor, and Doppler), together with measurements made with a commodity receiver worn by a person. This communication combines the two.

\begin{figure}[!t]
\centering
\includegraphics[width=\columnwidth]{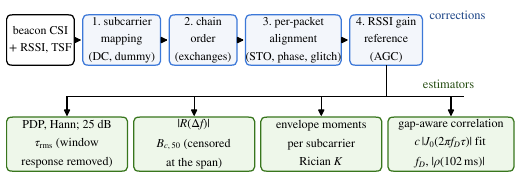}
\caption{Processing chain: four corrections, then four estimators; the delay-spread estimator also removes the window response (fifth correction).}
\label{fig:chain}
\end{figure}

\looseness=-1 Section~II gives the processing chain with five corrections, including one for unreported receive-chain exchanges (Fig.~\ref{fig:chain}); Section~III, its error budget from synthetic ground truth; and Section~IV, body-worn and handheld measurements with a capture-length guideline.

\section{Measurement Model and Processing Chain}
\subsection{CSI model}
For packet $n$, receive chain $r$, and subcarrier $k$, the CSI reported by a commodity receiver is
\begin{equation}
\hat h_{n,r}(k)=a_{n,r}\,e^{j(\phi_n+\psi_r)}\,e^{-j2\pi k\Delta f\,\tau_n}\,h_{n,\pi_n(r)}(k)+w_{n,r}(k),
\label{eq:model}
\end{equation}
\looseness=-1 where $h_{n,r}(k)$ is the channel to antenna $r$, $\pi_n$ an unreported assignment of antennas to chains, $a_{n,r}$ the gain set by automatic gain control (AGC), $\phi_n$ a common phase, $\psi_r$ a fixed chain phase, $\tau_n$ the symbol timing offset (STO), $\Delta f=312.5$~kHz (78.125~kHz for 802.11ax/be data frames, reported on a 156.25-kHz grid), and $w$ noise. The reported vector also carries a residual value at DC and implementation-specific padding.

\looseness=-1 We use beacons, which are broadcast non-high-throughput (non-HT) frames sent once per beacon interval of 102.4~ms and are therefore not beamformed. Their CSI covers the 53 subcarriers ($-26,\dots,26$) of the primary 20~MHz channel. Multi-stream data frames are often transmit-beamformed. Their CSI then estimates $\mathbf{H}(k)\mathbf{Q}(k)$ with a precoder $\mathbf{Q}(k)$ that follows the eigenvectors of $\mathbf{H}(k)$, which biases per-entry delay spread and coherence bandwidth \cite{companion}. The companion letter detects this blindly from the near-orthogonality and power ordering of the stream columns, with simulated false-alarm rates of at most $2.9\times10^{-3}$ and at least 98\% detection of fresh-feedback precoding at a signal-to-noise ratio (SNR) of 20--30~dB in i.i.d., line-of-sight (LOS), and moderately correlated channels \cite{companion}. The test classifies 279 of 291, 302 of 304, and 282 of 282 of the two-stream data frames used here (Section~IV-A) as precoded (AP-A at 80~MHz; AP-B at 80 and 160~MHz), and none of AP-C's 5221 in IEEE 802.11n mode or AP-D's 211 with beamforming off (the letter adds a second session). The total power is not biased: $\|\mathbf{H}\mathbf{Q}\|_F^2=\|\mathbf{H}\|_F^2$ for unitary $\mathbf{Q}$, and with four transmit antennas it stays within 1.7~dB (95th percentile) in simulation \cite{companion}. If an access point (AP) sent beacons from several chains with the cyclic shifts defined for the pre-HT fields \cite{ieee80211_2020}, an equal-power mixture would give $\tau_{\mathrm{rms,meas}}^2=\tau_{\mathrm{rms}}^2+\sigma_\delta^2$, where $\sigma_\delta$, the RMS of the shift set, is 100, 81.6, and 55.9~ns for the two- to four-chain sets $\{0,-200\}$, $\{0,-100,-200\}$, and $\{0,-50,-100,-150\}$~ns, and at least 55.9~ns for up to eight chains. Section~IV tests this.

\subsection{Corrections}
\emph{1) Subcarrier mapping.} The raw values are assigned to subcarrier indices. The DC residual and the padding are discarded, and DC is interpolated from its neighbors.

\looseness=-1 \emph{2) Chain order.} An exchanged packet looks like a burst of diffuse power and misleads the joint alignment. Each packet's per-chain amplitude and inter-chain phase patterns, which an exchange swaps and conjugates, are compared with those of the 60 nearest packets; runs that match better exchanged, with abrupt ends, are re-ordered with their received signal strength indicator (RSSI) if it follows. A capture is corrected only if this recovers exchanges inserted into 25\% of its packets with a balanced accuracy of at least 0.95.

\looseness=-1 \emph{3) Per-packet alignment.} The timing slope and common phase of every packet are estimated jointly across chains, first from the phase increment between adjacent subcarriers and then by matching each packet to the capture mean. This removes $\tau_n$ and $\phi_n$ in \eqref{eq:model} while keeping the inter-chain phase. The delay estimate is continuous, so whole-sample jumps of $\tau_n$ are removed like fractional ones; synthetic jumps of up to four 50-ns samples change $\tau_{\mathrm{rms}}$ by under 2\%. Isolated packets whose correlation with the capture mean is below 0.2 are discarded, but only when the capture is otherwise stable (median correlation above 0.8). Packets that place more than 5\% of their impulse-response energy more than 800~ns from the aligned delay centroid are discarded as corrupted (one of 28{,}413 beacons).

\emph{4) Gain reference.} Each packet and chain is scaled so that its mean subcarrier power follows the reported per-chain RSSI; packets without a valid RSSI are dropped (one beacon). This replaces $a_{n,r}$ with the measured received power.

\subsection{Estimators}
\looseness=-1 The power delay profile (PDP) is the average over packets and chains of $|\mathrm{IFFT}|^2$ of the Hann-windowed CSI, computed with a 256-point transform (12.5-ns bins over 3.2~$\mu$s) after alignment has placed the delay centroid at 400~ns, so that the whole response lies in the analyzed first half. $\tau_{\mathrm{meas}}$ is its second central moment above a threshold 25~dB below the peak. As the fifth correction, the window's own RMS width $\tau_w$, obtained by passing a single path through the same window, transform, placement, and threshold (35.3~ns for 53 subcarriers), is removed in quadrature, since second moments add under convolution:
\begin{equation}
\tau_{\mathrm{rms}}=\sqrt{\tau_{\mathrm{meas}}^2-\tau_w^2}.
\label{eq:quad}
\end{equation}
\looseness=-1 The threshold truncates both profiles, so \eqref{eq:quad} is approximate; the error budget of Section~III includes this effect. The coherence bandwidth $B_{c,50}$ is the lag at which the magnitude of the frequency autocorrelation, averaged over packets and chains, falls to 0.5. It is censored when it exceeds the 16.25~MHz span. $K$ is the median over subcarriers of the envelope-moment estimate \cite{greenstein1999moment,tepedelenlioglu2003ricean}, which uses magnitudes only and is therefore insensitive to residual phase. In synthetic tests it is within 0.6~dB from 0 to 12~dB and reads 0.5--1.1~dB low from 15 to 24~dB. The temporal autocorrelation of the zero-mean CSI pairs each packet with the packet nearest to $t+m\cdot102.4$~ms, within half an interval, so missed beacons do not stretch the lag axis. The maximum Doppler frequency $f_D$ comes from a fit of $c\,|J_0(2\pi f_D\tau)|$ \cite{clarke1968statistical}, where $c\le1$ absorbs decorrelation faster than the fitted $J_0$, including noise. For quasi-static links the fitted $f_D$ is an effective decorrelation rate of the diffuse part, not the Doppler frequency of moving scatterers. Confidence intervals use a moving-block bootstrap over packets (block length 5) \cite{kunsch1989jackknife}.

\begin{figure*}[!t]
\centering
\includegraphics[width=\textwidth]{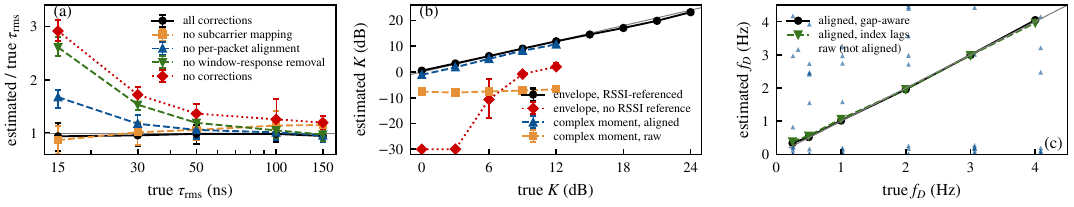}
\caption{Synthetic ground truth: medians and interquartile ranges. (a) Ratio of estimated to true $\tau_{\mathrm{rms}}$ with all corrections and with one or all removed (static channels, 150 realizations per point). (b) $K$ from envelope moments with and without the RSSI reference, and from the complex-mean estimator with and without alignment ($f_D=1$~Hz, 300 beacons, 20 realizations; above 12~dB, all corrections only). (c) Fitted $f_D$ with gap-aware and index-based lags, and from unaligned CSI (600 beacons, 10\% missed, 10 realizations).}
\label{fig:synth}
\end{figure*}

\begin{table}[!t]
\centering
\caption{Error Budget: Median Bias of $\tau_{\mathrm{rms}}$ (\%) and Ratio of Estimated to True $B_{c,50}$ (Static Channels, 150 Realizations per Point)}
\label{tab:budget}
\footnotesize
\setlength{\tabcolsep}{1.9pt}
\begin{tabular*}{\columnwidth}{@{\extracolsep{\fill}}lrrrrrcc@{}}
\toprule
 & \multicolumn{5}{c}{true $\tau_{\mathrm{rms}}$ (ns)} & \multicolumn{2}{c}{$B_{c,50}$ ratio}\\
\cmidrule(lr){2-6}\cmidrule(lr){7-8}
Processing & 15 & 30 & 50 & 100 & 150 & 30 ns & 50 ns\\
\midrule
All corrections & $-$6 & $-$4 & $-$1 & $-$2 & $-$5 & 1.00 & 1.00\\
No subcarrier map & $-$13 & +2 & +6 & +14 & +15 & 0.98 & 0.98\\
No chain-order fix$^{b}$ & +7 & +9 & +5 & +4 & $-$4 & 0.98 & 0.88\\
No alignment & +67 & +17 & +6 & +1 & $-$5 & 0.54 & 0.61\\
No window removal & +161 & +53 & +19 & +5 & $-$3 & 1.00 & 1.00\\
No RSSI reference & $-$6 & $-$4 & $-$1 & $-$1 & $-$6 & 1.00 & 1.00\\
No corrections & +191 & +72 & +36 & +26 & +20 & 0.50 & 0.57\\
\midrule
Fading, all corr.$^{a}$ & +14 & +10 & +13 & +4 & +1 & -- & --\\
\bottomrule
\end{tabular*}
\par\vspace{2pt}\parbox{\columnwidth}{\scriptsize $^{a}$Clarke fading on every tap, $f_D=1$~Hz, 200 beacons, 100 realizations; truth is the nominal PDP. $^{b}$25\% of beacons exchanged, correction~2 omitted; with it, the first row is restored, without a false exchange in 750 captures. Other rows: no exchanges.}
\end{table}

\begin{figure}[!t]
\centering
\includegraphics[width=\columnwidth]{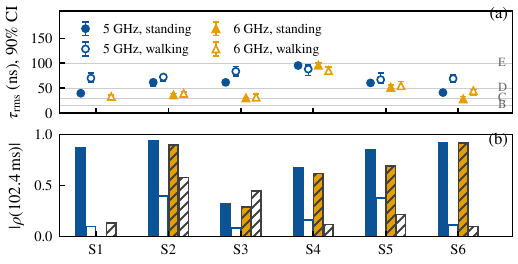}
\caption{Site survey. (a) $\tau_{\mathrm{rms}}$ with 90\% moving-block bootstrap intervals (within capture; drift between captures is larger, Section~IV-C); gray lines mark the RMS delay spreads of TGn models B--E. (b) Temporal correlation over one beacon interval (filled: standing; open: walking; hatched: 6~GHz).}
\label{fig:sites}
\end{figure}

\begin{figure}[!t]
\centering
\includegraphics[width=\columnwidth]{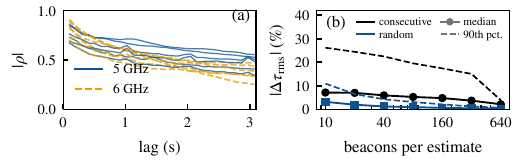}
\caption{Twelve 100-s captures of AP-B. (a) Gap-aware temporal autocorrelation. (b) Median and 90th-percentile deviation of $\tau_{\mathrm{rms}}$ from $N$ consecutive or randomly drawn beacons relative to the full capture.}
\label{fig:long}
\end{figure}

\section{Error Budget From Synthetic Ground Truth}
\looseness=-1 The generator produces two-chain CSI in the receiver's raw layout. It applies the impairments of \eqref{eq:model}: a uniform random $\phi_n$, $\tau_n\sim\mathcal{N}(150,20^2)$~ns, AGC steps with 2.7~dB standard deviation, a fixed chain phase, 30~dB SNR, a random DC residual and padding value, RSSI quantized to 1~dB, and 5--10\% missed beacons. Channels have exponential PDPs with 5~ns taps and $\tau_{\mathrm{rms}}$ from 15 to 150~ns (the span of TGn models B--F \cite{erceg2004tgn}). Some add a static line-of-sight path with Rician $K$, and some apply Clarke fading (moving receiver) to every tap \cite{clarke1968statistical}; the TGn spectrum (people moving at 1.2~km/h) \cite{erceg2004tgn} changes the fading row by at most 3 points. The truth is computed from the realized tap powers for static channels and from the nominal PDP for fading channels.

\looseness=-1 Table~\ref{tab:budget} and Fig.~\ref{fig:synth} give the result. Without alignment, random STO smears the averaged PDP and decorrelates the frequency response across packets. Without removing the window response, the 35~ns kernel dominates small spreads. Without subcarrier mapping, the misplaced grid and the DC residual bias large spreads. The RSSI reference does not affect $\tau_{\mathrm{rms}}$, but it decides $K$. Without it, AGC steps read as fading, and $K$ collapses to the $-30$~dB floor for true values of 0--3~dB and to $-10.7$~dB for 6~dB. Exchanged chains left in place raise small spreads by 13 points and lower $B_{c,50}$ (Table~\ref{tab:budget}). Without alignment, the correlation over one beacon interval is 0.03 for every $f_D$, so Doppler cannot be measured. For Clarke fading, the fitted $f_D$ of the full chain is within 0.1~Hz of the truth from 0.25 to 4~Hz; gap-aware lags matter only at small $f_D$. With all corrections, the median bias of $\tau_{\mathrm{rms}}$ is $-1\%$ to $-6\%$ for static channels and $+1\%$ to $+14\%$ for fading channels averaged over 200 beacons. A single static realization scatters widely around the median (10th to 90th percentile $\pm59\%$ at 15~ns, about $\pm20\%$ at 100--150~ns), because paths within one resolution cell (about 60~ns) interfere; only motion or wider bandwidth averages this out. Thresholds deeper than 25~dB raised measured 33-GHz road-channel spreads, converging near 40~dB \cite{wojtun2026sensitivity}; with 53 Hann-windowed subcarriers, beyond about 30~dB they admit the window's first sidelobe at $-31.5$~dB.

\looseness=-1 The budget is conditional on the impairment model, which has a flat receiver response. Averaged over the 35 captures of AP-B (Section~IV, two chains each), the measured power spectrum of each chain departs from flat by 3.3--3.4~dB (5~GHz) and 4.3--4.7~dB (6~GHz) peak to peak, mostly as one or two cycles across the band; its reproducible part, estimated from split halves of the captures, has a standard deviation of 0.9 and 1.2~dB. Applied to synthetic static channels in paired runs (300 per point), this response raises $\tau_{\mathrm{rms}}$ by 18--26\% at 15~ns, 7--11\% at 24~ns, 5--6\% at 30~ns, 3--4\% at 40~ns, and 2\% at 50~ns, a bias that the flat-response entries of Table~\ref{tab:budget} do not include. The measured AP-B spreads are 28~ns or more. Dividing the response, estimated from the other captures, out of the 13 captures at 6~GHz with $\tau_{\mathrm{rms}}\le45$~ns (median 33~ns) lowers $\tau_{\mathrm{rms}}$ by a median of 1.0\%, less than the modelled bias, because the division also imposes the error of the estimate: emulating the test in synthetic captures of the same spreads, with the smooth part of the measured response as truth and an estimate formed in the same way, gives a first-application bias of 4.8--5.5\% and a divide-out change of $-1.1\%$ to $-2.4\%$ (medians of 300 trials, diffuse or 6-dB LOS). Over all AP-B captures (up to 96~ns), the measured median change is $-0.1\%$ (5~GHz) and $-1.0\%$ (6~GHz), at most 7.4\%.

\begin{table*}[!t]
\centering
\caption{Site Survey With the Body-Worn Receiver (AP-B, Primary 20~MHz; Corrections of Section~II Applied)}
\label{tab:meas}
\scriptsize
\setlength{\tabcolsep}{4.2pt}
\begin{tabular*}{\textwidth}{@{\extracolsep{\fill}}lcccccccccccccc@{}}
\toprule
 & \multicolumn{4}{c}{$\tau_{\mathrm{rms}}$ (ns)} & \multicolumn{2}{c}{$B_{c,50}$ (MHz), standing} & \multicolumn{4}{c}{$K$ (dB)} & \multicolumn{4}{c}{$|\rho(102.4\,\mathrm{ms})|$}\\
\cmidrule(lr){2-5}\cmidrule(lr){6-7}\cmidrule(lr){8-11}\cmidrule(lr){12-15}
 & \multicolumn{2}{c}{standing} & \multicolumn{2}{c}{walking} & & & \multicolumn{2}{c}{standing} & \multicolumn{2}{c}{walking} & \multicolumn{2}{c}{5 GHz} & \multicolumn{2}{c}{6 GHz}\\
\cmidrule(lr){2-3}\cmidrule(lr){4-5}\cmidrule(lr){8-9}\cmidrule(lr){10-11}\cmidrule(lr){12-13}\cmidrule(lr){14-15}
Site & 5 GHz & 6 GHz & 5 GHz & 6 GHz & 5 GHz & 6 GHz & 5 GHz & 6 GHz & 5 GHz & 6 GHz & stand. & walk. & stand. & walk.\\
\midrule
S1 & 40 & -- & 70 & 33 & $\ge$16 & -- & 14 & -- & $<\!-10$ & $-7$ & 0.87 & 0.10 & -- & 0.13\\
S2 & 62 & 36 & 72 & 39 & 15.3 & 6.4 & 8 & 12 & $<\!-10$ & 1 & 0.93 & 0.40 & 0.90 & 0.57\\
S3 & 62 & 30 & 84 & 32 & 3.5 & 7.1 & 6 & 5 & $<\!-10$ & $<\!-10$ & 0.32 & 0.08 & 0.29 & 0.45\\
S4 & 96 & 96 & 88 & 85 & 3.6 & 2.6 & 7 & 5 & $<\!-10$ & $<\!-10$ & 0.67 & 0.16 & 0.62 & 0.11\\
S5 & 61 & 51 & 67 & 55 & $\ge$16 & 3.9 & 13 & 6 & $<\!-10$ & $<\!-10$ & 0.85 & 0.38 & 0.69 & 0.22\\
S6 & 41 & 28 & 69 & 45 & 12.1 & 9.7 & 11 & 6 & $<\!-10$ & $-2$ & 0.92 & 0.11 & 0.91 & 0.10\\
\bottomrule
\end{tabular*}
\par\vspace{2pt}\parbox{\textwidth}{\scriptsize $\ge$16: censored at the 16.25~MHz span. --: no beacons in that band. $<\!-10$: no static component. 90\% intervals of $\tau_{\mathrm{rms}}$: Fig.~\ref{fig:sites}(a). Text values use unrounded data.}
\end{table*}

\begin{table*}[!t]
\centering
\caption{Long Captures, Further Access Points, and Wideband Data Frames}
\label{tab:sources}
\scriptsize
\setlength{\tabcolsep}{4.0pt}
\begin{tabular*}{\textwidth}{@{\extracolsep{\fill}}lcccccccccc@{}}
\toprule
Source & Frames & BW (MHz) & $n$ & RSSI (dBm) & $\tau_{\mathrm{rms}}$ (ns) & $B_{c,50}$ (MHz) & $K$ (dB) & $\sigma_P$ (dB) & 10\% fade (dB) & Pattern corr.\\
\midrule
AP-B, 5 GHz, 6$\times$100 s$^{a}$ & B & 20 & 5628 & $-40$ & 42--57 & 3.7--$\ge$16 & 7--11 & 2.2 & 3.4 & 0.9917\\
AP-B, 6 GHz, 6$\times$100 s$^{a}$ & B & 20 & 5622 & $-56$ & 28--42 & 4.1--$\ge$16 & 4--10 & 2.4 & 3.8 & 0.9873\\
AP-A, 5 GHz & B & 20 & 884 & $-40$ & 50 & 8.8 & 10 & 2.0 & 1.7 & 0.9993\\
AP-C, 5 GHz, 36$\times$30 s$^{a}$ & B & 20 & 9778 & $-37$ & 58--88 & 2.0--3.9 & $<\!-10$ to 6 & 1.8--4.1 & 2.4--6.2 & 0.01--0.99\\
AP-D, 5 GHz, 8$\times$30 s$^{a}$ & B & 20 & 1772 & $-42$ & 53--87 & 2.1--3.1 & $<\!-10$ to 7 & 1.6--3.3 & 2.5--5.2 & 0.18--0.97\\
AP-E, 5 GHz, 2 captures$^{a}$ & B & 20 & 2279 & $-33$ & 24--25 & 8.6 & 8--17$^{c}$ & 4.5--4.6 & 9.7 & 0.9998--0.9999\\
AP-F, 5 GHz & B & 20 & 1025 & $-70$ & 61 & 3.0 & 8$^{c}$ & 3.0 & 4.0 & 0.9512\\
AP-G, 5 GHz & B & 20 & 585 & $-25$ & 64 & 3.1 & 19$^{c}$ & 4.4 & 6.1 & 0.9993\\
\midrule
AP-A, 5 GHz & D & 80 & 291 & -- & -- & -- & -- & 1.9 & 1.5 & 0.9986\\
AP-B, 5 GHz$^{b}$ & D & 80 & 304 & -- & -- & -- & -- & 2.0 & 3.1 & 0.9445\\
AP-B, 6 GHz$^{b}$ & D & 160 & 282 & -- & -- & -- & -- & 3.0 & 3.2 & 0.9661\\
\bottomrule
\end{tabular*}
\par\vspace{2pt}\parbox{\textwidth}{\scriptsize Frames: B, beacons (20~MHz primary channel); D, two-stream data frames (precoded). $\sigma_P$: standard deviation over subcarriers of $10\log_{10}P(k)$, median over frames. 10\% fade: median over frames of the median minus the 10th percentile of $10\log_{10}P(k)$. Pattern corr.: median correlation of $10\log_{10}P(k)$ between consecutive frames (beacons 0.10~s apart, AP-A data 0.28~s, AP-B data 1.0~s). $^{a}$Ranges over the captures; RSSI is their median. $<\!-10$: no measurable static component. $^{b}$Second client and session. $^{c}$Lower bound (Section~IV-D).}
\end{table*}

\section{Measurements}
\subsection{Sources}
\looseness=-1 Eight capture sets with seven access points (AP-A to AP-G; AP-B in three sets, AP-C in two modes) span environments, motion, bands, bandwidths, and frame types. Every receiver was a commercial two-chain Wi-Fi client reporting CSI, per-chain RSSI, and the timing synchronization function (TSF) timestamp per frame.

\looseness=-1 \emph{Site survey.} A wearable device with such a client (two antennas about 10~cm apart), worn on the head by a person, recorded beacons from an IEEE 802.11be AP (AP-B) on primary channels at 5745~MHz (U-NII-3) and 6135~MHz (U-NII-5). There were six sites, each 5--10~m from the AP, with line of sight except in the corridors: two laboratories (S1, S2), a conference room (S3), two corridors (S4, S5), and a lobby (S6). At each site the person stood still or walked at a normal pace, giving 23 captures of 24--49 beacons (2.4--4.9~s).

\emph{Long captures.} Twelve captures of AP-B, six per band, each 98--102~s and 909--990 beacons, at a second position in S1, worn by a person standing 2--3~m from the AP.

\emph{AP-A.} A 100-s beacon capture of an IEEE 802.11ax AP (AP-A) in the U-NII-1 band, with 291 two-stream data frames from it in the same session at S1, the client on a desk 2--3~m from the AP.

\looseness=-1 \emph{One AP in two modes.} A second device with the same chipset, held or carried by one of two persons standing, walking, and walking fast, recorded 30-s captures of AP-C's beacons and two-stream data frames (ICMP echo, ten per second) at S2, a conference room, and an outdoor area, with AP-C in IEEE 802.11n mode (40~MHz, unprecoded) and in 802.11ac mode (80~MHz, precoded \cite{companion}): 36 beacon captures.

\looseness=-1 \emph{AP with switchable beamforming.} A device with the same chipset recorded data frames (80~MHz at 5~GHz) and beacons of AP-D, a $4\times4$ IEEE 802.11ax AP, with transmit beamforming off and on, about 30~s each, on a table, held by a standing person, or carried at two walking speeds (beacon RSSI $-38$ to $-44$~dBm), and in a third session eight 10-s captures of the same conditions with data frames at 300--345 per second (ICMP echo every 1~ms; 25{,}909 frames).

\emph{Further APs.} Beacon captures of three more access points in the U-NII-3 band: AP-E (two captures), AP-F, and AP-G. They last 60--180~s, with mean RSSI from $-70$ to $-25$~dBm, in a third laboratory.

\looseness=-1 \emph{Wideband data frames.} Two-stream IEEE 802.11be data frames from AP-B at 80~MHz (5~GHz) and 160~MHz (6~GHz), one per second (ICMP echo), received by a second client held 5--10~m from the AP in the third laboratory. These are the frames of the six 100-s captures with ICMP traffic, less four 80-MHz frames with zero-power subcarriers (the letter keeps the four and adds 14 at 80~MHz and one at 160~MHz from captures without ICMP traffic).

\looseness=-1 Only the longest segment of each capture without a gap above 10~s was used. As the wideband data frames are precoded, we report only statistics of their total power $P(k)=\sum_{r,s}|\hat h_{r,s}(k)|^2$, which precoding nearly preserves \cite{companion}. Frame by frame, the 1-ms captures show the client exchanging its chains for one or two beacon intervals (28--29\% of the frames on the table, 18\% held with beamforming on, 1--2\% held or walking with it off); at the beacon rate, correction~2 restored both table captures and the held one with beamforming off (57 of 294 frames, taken one per beacon interval, exchanged) without error (0 of 57 missed, 0 of 237 falsely exchanged), which bounds misses below 5.2\% and false exchanges below 1.3\% (one-sided 95\%), and declined the other five, in which the per-chain patterns changed between those frames (walking, or precoder updates when held with beamforming on) and it would have missed most exchanges. Of the 84 beacon captures, 30 were determinable (none walking), 19 of them with exchanges, from one beacon to 47\% of a capture; restoring them raised the site survey's standing median $|\rho(102.4\,\mathrm{ms})|$ from 0.74 to 0.85 and $K$ at AP-D's table from 2--3 to 6--7~dB, and lowered one long capture's $\tau_{\mathrm{rms}}$ from 67 to 48~ns.

\subsection{Site survey}
\looseness=-1 Table~\ref{tab:meas} and Fig.~\ref{fig:sites} summarize the 23 captures. The delay spreads range from 28~ns (S6, 6~GHz) to 96~ns at S4, which reads 85--96~ns in every condition; changing the threshold to 20 or 30~dB moves every site value by at most 10\%. The values lie roughly between the delay spreads of TGn models C and E, but this label describes $\tau_{\mathrm{rms}}$ only. For an exponential PDP, $B_{c,50}\tau_{\mathrm{rms}}=\sqrt3/(2\pi)\approx0.28$. Of the 84 beacon captures, 70 give 0.16--0.36, but 14 give 0.43 to above 0.98 (ten censored), and 12 of these have $K\ge4.5$~dB: a dominant path keeps the band correlated while weaker late paths set $\tau_{\mathrm{rms}}$. Standing captures show a static component with $K$ of 5--14~dB and a median correlation of 0.85 over one beacon interval. Walking largely removes the static component: $K$ is at most 1.2~dB in all 12 walking captures and below $-10$~dB in 9 of them, and the median correlation falls to 0.15, near the estimator's floor (from 0.09--0.11 at 24 to 0.06--0.07 at 49 walking beacons in shuffled order). At 5~GHz, walking increased $\tau_{\mathrm{rms}}$ at five of six sites, by up to 30~ns (S1, S6), consistent with the body intermittently shadowing the dominant path \cite{cotton2014statistical}. At three of the five sites with both bands (S2, S3, S6), the standing 6~GHz spread was 31--51\% below the 5~GHz value, more than the 23\% (90th percentile) drift of a short capture (Section~IV-C); at S5 it was 15\% lower and at S4 equal. The long captures of the same AP show the same ordering (42--57~ns at 5~GHz against 28--42~ns at 6~GHz, Table~\ref{tab:sources}). With five sites, one AP, and one wearer, this ordering is suggestive only; penetration loss and the AP's 6-GHz radio and antenna are candidate causes these data cannot separate. The smallest spreads, 28--30~ns, are well below the 55.9~ns or more that the standard's equal-power cyclic shifts for two to eight chains would add \cite{ieee80211_2020}, so AP-B does not send its beacons from several equal-power chains.

\subsection{Long captures and capture length}
\looseness=-1 The twelve long captures (Fig.~\ref{fig:long}) have $K$ of 3.8--10.9~dB and correlations of 0.67--0.91 at 102.4~ms, so 9--33\% of the diffuse power decorrelates within one beacon interval, consistent with small body movements and noise, and the rest over seconds, at a fitted effective rate of 0.06--0.10~Hz. For AP-B, forty consecutive beacons (4.1~s) differ from the 100-s value by a median of 6\% (90th percentile 23\%, 275 non-overlapping windows), whereas forty drawn at random from the whole capture differ by a median of 1\% (90th percentile 4\%). A short capture measures the channel at that moment; a survey should spread its beacons over time or positions and report the scatter.

\subsection{Further access points and wideband frames}
\looseness=-1 Across all seven APs (Tables~\ref{tab:meas} and~\ref{tab:sources}), $\tau_{\mathrm{rms}}$ ranges from 24 to 96~ns and $\sigma_P$ from 1.6 to 4.6~dB. At AP-E to AP-G, $|\rho(102.4\,\mathrm{ms})|$ was 0.56--0.67, and as estimation noise counts as diffuse power, their $K$ values are lower bounds. Walking lowered the pattern correlation (Table~\ref{tab:sources}) to 0.01--0.21 at AP-C and 0.18--0.49 at AP-D, against 0.67--0.99 and 0.83--0.97 standing or on a table. Over AP-C's 36 captures, walking lowered the median $|\rho(102.4\,\mathrm{ms})|$ from 0.65 to 0.05 (shuffled-order floor 0.02--0.03) for both persons at all three sites; for the same site and motion, $\tau_{\mathrm{rms}}$ differed by a median of 9\% between persons (at most 32\%) and 9\% between the two modes (at most 25\%), outside the 90\% intervals in 7 and 10 of 18 pairs, whereas, on the same 40~MHz, its data streams' $\tau_{\mathrm{rms}}$ in 802.11ac mode was 0.22--0.83 times that in 802.11n mode \cite{companion}. On the primary 20~MHz, AP-A's precoded data frames have a $\sigma_P$ of 0.4~dB (1.9~dB over 80~MHz), against 2.0~dB for its beacons. A data frame's $P(k)$ sums the channel over every transmit antenna, whereas a beacon's is a single coherent channel, so the two should not be pooled: at AP-C in 802.11n mode, without precoding, one data stream gives about the beacons' $\sigma_P$ (medians 3.4 and 3.1~dB), the sum over both streams 1.7~dB. Delay spreads differ too: through the same chain and on the beacon subcarriers, per-stream $\tau_{\mathrm{rms}}$ of unprecoded data frames was 0.64--1.21 times the beacon value at AP-C and 0.22--0.70 at AP-D, whose unbeamformed beacons changed by only $-8$ to $+9\%$ with beamforming on. By Section~II, the data streams' 18--37~ns plus the cyclic shifts of AP-D's four chains give 59--67~ns at equal power and up to 84~ns with unequal chain gains; only the slow walk's beacons, recorded 1.4~min before its data, exceed this range (by 3--12~ns). 

\subsection{Limitations}
\looseness=-1 The receivers were not compared with a reference sounder. Accuracy rests on the synthetic error budget, which assumes a flat receiver response and no quadrature imbalance, and on internal consistency: bootstrap intervals, agreement between short and long captures, the cyclic-shift check, and the measured static response. Beacon spreads can include an AP's transmit cyclic shifts; Section~IV-B excludes only equal-power ones. Exchanges are restored only where determinable, never in walking beacon captures, where a 25\% rate, within the measured range (Section~IV-A), changed $\tau_{\mathrm{rms}}$ by $-16$ to $+28\%$, $|\rho(102.4\,\mathrm{ms})|$ by up to 0.2 (site-survey walking median 0.19, not 0.15), and $K$ above $-10$~dB by at most 2~dB. The primary 20~MHz channel limits resolution to the 35-ns window response, removed but not undone; spreads below about 15~ns are biased. At one beacon every 102.4~ms, Doppler is resolvable only below 4.9~Hz, against 19--27~Hz for walking at 5.8~GHz, so walking is characterized by decorrelation, and the Doppler accuracy of Section~III is synthetic. The body-worn results come from one wearer and mounting position, and all receivers share one chipset family, whose CSI conventions underlie every result.

\section{Conclusion}
\looseness=-1 A commodity Wi-Fi receiver can serve as a device-referred channel sounder if its CSI is mapped to subcarriers, kept in receive-chain order, aligned per packet, and referenced to RSSI, and if the window response is removed; otherwise, delay spread, coherence bandwidth, $K$, and Doppler are biased by more than typical differences between sites. With them, a body-worn receiver resolved site and motion differences at 5 and 6~GHz from captures of a few seconds, and the chain, applied unchanged to beacons of seven APs, gave delay spreads of 24--96~ns. Beacon and data-frame CSI must not be pooled without knowing the AP's transmit processing: precoding alters data frames, and at one AP the beacons gave 1.4--4.5 times the unprecoded data streams' delay spread, as transmit cyclic shifts predict.

\section*{Acknowledgment}
The authors used Claude Opus 5.5 (Anthropic, Nov. 2026) to help draft and edit the text and to write the analysis and figure code; the authors checked all results, equations, and references.

\bibliographystyle{IEEEtran}
\bibliography{refs}

\end{document}